\documentclass[preprint,showpacs,pre,preprintnumbers,amsmath,amssymb]{revtex4}

\usepackage{graphicx}
\usepackage{dcolumn}
\usepackage{bm}

\begin{document}
\title{Fluctuating magnetic moments in liquid metals}
\author{Mark Patty, Keary Schoen and Wouter Montfrooij}
\affiliation{Department of Physics and Astronomy, and Missouri Research Reactor,
University of Missouri, Columbia, MO 65211}

\begin{abstract}
{We re-analyze literature data on neutron scattering by liquid metals to show that non-magnetic
liquid metals possess a magnetic moment that fluctuates on a picosecond time scale. This time scale
follows the motion of the cage-diffusion process in which an ion rattles around in the cage formed
by its neighbors. We find that these fluctuating magnetic moments are present in liquid Hg, Al, Ga
and Pb, and possibly also in the alkali metals.}
\end{abstract}
\pacs{61.12.-q, 61.25.Mv, 72.15.Cz}
\maketitle

\section{Introduction}
During the past decades the properties of a range of elemental liquid metals have been studied by
means of neutron scattering
\cite{liviaprl,badyal,bove,cs,bove3,k,k2,bove2,ru,page,na3,alblas,stangl,balucani2,torcini,burkel,dejong,cocking,north,soderstrom,reijers,iqbal,bf,fredrikze,egelstaff}
and X-ray scattering experiments\cite{tamura,tamura2,huijben,pilgrim,scopigno,li3,dzuxray,xray}.
Unlike ordinary simple fluids, liquid metals can support short-wavelength sound waves far outside
the hydrodynamic regime; simple fluids only support very strongly damped density fluctuations beyond
the hydrodynamic region\cite{host}. In other words, a density disturbance decays much faster in a
simple fluid than it does in a liquid metal under comparable thermodynamic conditions. Typically,
a short-wavelength sound wave in a simple liquid does not propagate beyond one wavelength\cite{host}.
Presumably, this difference can be attributed to the presence of two interacting systems in a liquid
metal: the positively charged ionic liquid and the negatively charged conduction sea. This notion has
stimulated the study of the decay mechanism of the density fluctuations, by means of neutron and X-ray
scattering experiments, as well as by molecular dynamics (MD) computer simulations in a range of liquid
metals, such as Hg\cite{liviaprl,badyal,bove,tamura,tamura2}, Cs\cite{cs,bove3,huijben}, K\cite{k,k2,bove2},
Rb\cite{ru,page}, Na\cite{na3,alblas,stangl,balucani2,pilgrim,scopigno},
Li\cite{torcini,burkel,dejong,li3}, Pb\cite{cocking,north,soderstrom,reijers,dzuxray},
Al\cite{iqbal,xray}, and Ga\cite{bf}. These studies have by and large confirmed the role
of the electron sea as a feedback mechanism, serving to reduce the decay rates of disturbances,
and also ensuring that density fluctuations can propagate at a higher velocity than the adiabatic
sound velocity.\\

These studies also showed that similar to ordinary liquids, cage
diffusion plays an important part in the decay mechanism of
density fluctuations\cite{host,cage,peter,pusey,schepper,bove}.
Cage diffusion occurs when an atom bounces off neighboring atoms,
thereby confining the atom to a "cage." This is in contrast to
self-diffusion, the process in which the atom moves through the
sample and which is characterized by a net displacement from its
starting position over a period of time\cite{host}. In MD
simulations, where one follows the position of an atom over time,
cage-diffusion and self-diffusion show up as two distinct time
scales\cite{liviaprl}. Cage-diffusion accounts for a small
decrease in correlation between the initial and subsequent
position of an atom; this initial decrease in correlation occurs
within a few picoseconds. The overall demise of correlation is
given by the self-diffusion process, which takes place on a much
longer time scale\cite{host} and is determined by the coefficient
for self-diffusion $D_s$. These two diffusive processes can also
be observed by means of quasi-elastic neutron
scattering\cite{badyal}. Neutron scattering is sensitive to the
motion of individual atoms because an atom moves during the time
it takes the neutron to interact with it\cite{squires}. This
motion shows itself as a spread in energy of the scattered neutron
wave-packet. Rapid movement (cage-diffusion) results in a large
spread in energy; slow movement (self-diffusion) results in a
spread with small characteristic energy-width. Both these
processes have indeed been observed in liquid metals. For
instance, in liquid mercury\cite{liviaprl,badyal,bove}, the
scattered neutron intensity originating from a single atom (the
so-called incoherent scattering contribution\cite{squires})
corresponds to a superposition of two Lorentzian lines. One line
is sharp (in energy), corresponding to self-diffusion, and one
line is broad, corresponding to cage-diffusion. A Lorentzian line
in energy corresponds to an exponential decay in
time\cite{squires} of the correlation
between the initial and subsequent position of an atom.\\

A comparison between the neutron scattering data and the MD
simulations on liquid Hg revealed a serious discrepancy regarding
the effectiveness of the cage-diffusion
mechanism\cite{liviaprl,badyal,bove}. While both studies agreed on
the characteristic time scale for the cage-diffusion process,
according to the neutron scattering study\cite{badyal} cage
diffusion accounted for up to 20\% of the loss in correlation in
the position of an atom, compared to only 0.4\% as observed in the
MD results\cite{bove}. In order to explain this discrepancy,
Badyal {\it et al.}\cite{badyal} suggested that a mercury ion
might have a fluctuating magnetic moment, resulting in an enhanced
neutron scattering cross-section. The idea here is straightforward
(see Fig. 1): in a liquid, atoms can approach each other very
closely. On such a close approach, an electron from a filled inner
shell of the metallic ion can be ejected into the Fermi-sea (Fig.
1b), resulting in an unpaired electron, and hence in a magnetic
moment (Fig. 1c). Once the ions move away from each other again,
the shell can be re-completed (Fig. 1d). One can thus expect a
magnetic moment to pop in and out of existence on the same time
scale as the rattling motion of an atom inside its cage. This
process automatically leads to a pathway for the neutron to
scatter from the atom via the electromagnetic force\cite{squires},
augmenting the interaction via the strong nuclear force and
resulting in an enhanced cross section for the cage-diffusion
process. From the strength of the magnetic
interaction\cite{squires}, it can then be determined what fraction
of the time an ion has an
unpaired electron.\\

In this paper we show that the cage-diffusion process in liquid metals is indeed accompanied by a
fluctuating magnetic moment. We do this by revisiting published neutron scattering data on Hg, Cs, K, Rb,
Na, Li, Pb, Al, and Ga. We observe a small effect in the alkali metals, but find that the ions in Ga
and Hg have unpaired electrons for up to 20\% of the time. Not only do these magnetic moments provide
an additional means for studying cage-diffusion by means of neutron scattering, they provide an additional
long-range interaction mechanism for the ions in the liquid.\\

\section{Theory}

In this section we briefly review the various contributions that make up the neutron scattering cross
section of a liquid. We use the data by Badyal {\it et al.}\cite{badyal} on mercury to illustrate the
various contributions, and to demonstrate under what conditions one can observe the proposed fluctuating
magnetic moments.\\
A neutron interacts with the nucleus of an atom via the strong
nuclear force, and with the magnetic moments of electrons present
in the system via the electromagnetic force\cite{squires}. Thus,
the total number of neutrons with initial energy $E_i$ that are
scattered every second into a solid angle d$\Omega$ having final
energies between $E_f$ and $E_f +dE$ is given by the double
differential cross section and can be separated into a nuclear and
a magnetic term\cite{squires}:

\begin{equation}
\frac{d^2\sigma_{total}}{d\Omega dE}=\frac{d^2\sigma_{nuclear}}{d\Omega dE}+
\frac{d^2\sigma_{magnetic}}{d\Omega dE}.
\end{equation}
For mono-atomic systems, such as the ones considered in this paper, the nuclear contribution for single
scattering events is given by
\begin{equation}
\frac{d^2\sigma_{nuclear}}{d\Omega dE}=\frac{k_f}{k_i}\frac{\sigma_{coh}}{4\pi}S_{coh}(q,E)+\frac{k_f}{k_i}\frac{\sigma_{inc}}{4\pi}S_{inc}(q,E)
\label{main}
\end{equation}
$S_{coh}(q,E)$ is the dynamic structure factor and represents the collective response of the liquid as
a function of momentum $\hbar q$ and energy E transferred from the neutron to the liquid, while
$S_{inc}(q,E)$ describes the dynamics of a single atom\cite{squires}. The cross-sections $\sigma_{coh}$
and $\sigma_{inc}$ are element dependent; $\sigma_{inc}$ arises because the strong interaction depends
on the spin state of the nucleus and the number of neutrons in the nucleus. Thus, the nuclear scattering
cross-section carries information about the collective behavior of the atoms, such as soundwaves, and
information about the motion of indiviual atoms, such as self-diffusion.\\

The static structure factor $S(q)$ is given by the sum-rule\cite{squires}
\begin{equation}
S(q)= \int  S_{coh}(q,E)dE,
\label{sq}
\end{equation}
while the incoherent dynamic structure factor satisfies a similar sum-rule
\begin{equation}
1= \int S_{inc}(q,E)dE.
\end{equation}
In neutron diffraction experiments aimed at measuring $S(q)$, the energy integration in Eq. \ref{sq} is
carried out by the neutron detector. Because of the term $k_f/k_i$ in Eq. \ref{main}, this procedure leads
to small errors in the determination of $S(q)$; however, these errors are small under suitable experimental
conditions and can be corrected for using standard methods\cite{squires}. A further source of errors is
that Eq. \ref{main} is only valid for neutrons that are scattered once by the sample. Again, by choosing
sufficiently small samples, the errors introduced by multiple scattering events and events in which a
scattered neutron is absorbed by the sample can be corrected for\cite{sears}. Therefore, provided these
corrections have been carried out, one can check the accuracy of the data reduction procedure by comparing
the measured cross-sections $\sigma_{coh}$ and $\sigma_{inc}$ to the known values. Even in the case where
the absolute values of $\sigma_{coh}$ and $\sigma_{inc}$ cannot be inferred from the experiment, their
ratio can still be determined using the fact that $S(q)$ oscillates around 1 for large $q$.\\

The magnetic contribution to the scattered intensity is only
visible in neutron scattering experiments on liquids provided that
atoms with unpaired electrons exist\cite{squires}. The angular
momentum associated with these unpaired electrons, h$J$, interacts
with the intrinsic magnetic moment of the neutron. The conduction
electrons present in liquid metals do not contribute to the
scattering at finite $q$; an electron moves so fast compared to
the neutron that the scattered waves only add up coherently at
$q$=0, the forward direction. However, if an electron is localized
around an atom, all scattered waves originate from the region of
the partially filled orbital, and the scattered waves can be
observed for a range of q-values. For this reason the form factor
for magnetic scattering $F(q)$, which describes the variation of
scattered intensity with $q$ and which is given by the spatial
extent of the electron cloud, falls off more rapidly with
increasing $q$ than the form factor for nuclear scattering [the
so-called Debye-Waller factor $W(q)$]. The latter reflects the
fact that nuclear scattering originates in the much smaller volume of the nucleus.\\

The number of ions with unpaired electrons at any given moment determines the magnetic cross-section for a
liquid with fluctuating magnetic moments. The total number of neutrons that are scattered per second per
metallic ion into solid angle d$\Omega$ is given by the paramagnetic approximation for the differential
cross section\cite{squires}
\begin{equation}
\frac{d\sigma^{magnetic}}{d\Omega}=n\frac{2}{3}(\gamma r_0)^2[\frac{1}{2}g(LJS)F(q)]^2e^{-2W(q)}J(J+1).
\label{para}
\end{equation}
In this equation, $n$ is the fraction of the ions that have a
collision induced angular momentum h$J$,
$g(LJS)=3/2+[S(S+1)-L(L+1)]/[2J(J+1)]$ describes how the intrinsic
angular momentum of the electron h$S$ and its orbital angular
momentum h$L$ add up to the magnetic moment $\mu_B g(LJS) J$
($\mu_B$ is one Bohr magneton), and $(\gamma r_0)^2$=0.291 barn is
the strength of the interaction with the neutron. Eq. \ref{para}
offers a good approximation of the strength of the magnetic
scattering provided that the characteristic energy width of the
quasi-elastic scattering as determined by the underlying
cage-diffusion mechanism is small compared to the incident energy
of the neutron\cite{squires}. This is the same requirement that
allows one to determine $S(q)$ from a liquid without doing an
energy analysis of the scattered neutron, and we will therefore
assume that this requirement is satisfied for all published
datasets discussed in this paper.\\

Analyzing quasi-elastic neutron scattering experiments on liquid
Hg at room temperature, Badyal {\it et al.}\cite{badyal} observed
that the scattered signal at small momentum transfers consisted of
two contributions (see Fig. \ref{hg}), attributable to
self-diffusion and cage-diffusion, respectively. However, the
relative strength (area under the curves in Fig. \ref{hg}) of the
cage-diffusion contribution compared to the self-diffusion
contribution was found to be 22\% (corresponding to a differential
cross section of $\sim$ 1.5/4$\pi$ barn). A relative strength of
the order of 0.3\% was expected based on MD simulations\cite{bove}
and on an order of magnitude calculation\cite{badyal}. Given that
the strength of the quasi-elastic coherent contribution for small
q-values (given by sum-rules at $\sim$0.01/4$\pi$ barn) was
negligible\cite{handbook,squires}, and given that the
characteristic energy width (3 ps$^{-1}$) corresponded to the time
scale of the cage diffusion process (1/3 ps), the
authors\cite{badyal} concluded that the broad quasi-elastic line
did indeed correspond to cage-diffusion but with a magnetically
enhanced cross-section. Using Eq. \ref{para} ($S$= 1/2, $L$=2 and
$J$=5/2 and $F(q)$=e$^{-W(q)}$= 1 for small $q$) and noting that
crystal electric field effects are absent in a liquid, we find
that 19.5\% of the Hg-ions have an unpaired d-electron. Should the
observed magnetic signal originate from an unpaired s-electron,
then the corresponding fraction of magnetic ions would
be 82\%. We return to this latter possibility in the discussion.\\

Thus, a significant fraction of the mercury ions has a magnetic
moment; this moment can interact with its neighbors via the
magnetic dipole interaction, via the direct exchange interaction,
and via polarization of the conduction electrons. The dipole
interaction likely only adds up to a small correction to the
interatomic potential at small distances, but it becomes the
dominant interaction mechanism at large distances and therefore it
might well contribute to the ability of a liquid metal to sustain
propagating soundwaves with short wavelengths. Likewise, the
polarization of the conduction electrons by the atomic magnetic
moments provides a direct interaction mechanism between the ionic
liquid and the conduction electrons. It is the presence of the two
interacting systems that is presumably responsible for the
existence of well-defined short wavelength sound waves. For this
reason, we have re-analyzed existing neutron scattering data
\cite{cs,k,torcini,reijers,iqbal,bf} on liquid metals in order to
investigate the presence of magnetic moments in non-magnetic
liquids. We note that short-lived magnetic moments do not
contradict the overall diamagnetic response of a liquid metal:
macroscopic
measurements take place on a much larger time scale than the lifetime of a collision-induced atomic moment.\\

Fluctuating magnetic moments can betray their presence in various ways in neutron scattering experiments.
In diffraction experiments the additional cross-section would lead to an increased signal at
smaller q-values, decaying with $q$ according to $|F(q)|^2$. This additional signal would be on top of
the angle independent incoherent cross-section and the weakly angle dependent multiple scattering
cross-section. Thus, whether the proposed signal is actually visible in published data depends on
the strength of the incoherent cross section and on the details of the data reduction procedure.
It is easiest to identify the magnetic cross section in quasi-elastic neutron scattering experiments
 (as in the liquid mercury experiments\cite{badyal}); however, we did not find data sets in the
 literature suited to the latter approach. Finally, it is unclear a priori how an increase in
 temperature and density would affect the magnetic cross section. This increase would allow for closer
 approach of the ions thereby increasing the overlap of the filled orbitals; however, the life-times
 of the induced moments would likely decrease as well resulting in a signal that would be too spread
 out in energy to be reliably observable in neutron scattering experiments.\\

\section{Results}

Our investigation is limited to published studies that show the
raw data and detail the correction procedure, or to studies where
the incoherent scattering contribution is absent. Surprisingly,
this leaves very few data sets on liquid metals. In most
investigations the data are only presented after subtraction of
the contribution identified as incoherent scattering. This
subtraction procedure would also have eliminated the magnetic
contribution, should it have been present. Evaluation of the
published neutron scattering data on the much studied alkali
metals shows that the percentage of ions having a magnetic moment
is likely to be much smaller than what was observed in liquid
mercury, and that in most cases it is not possible to come to an
unambiguous conclusion whether this magnetic contribution is
present or not. On the other hand, the group 3 and 4 metals Al, Pb
and Ga show a large effect similar to
liquid mercury. All results are collected in Table I.\\

\subsection{The alkali metals}

Bodensteiner et al.\cite{cs} observed a discrepancy between the
value for the incoherent scattering cross-section as measured in
their inelastic neutron scattering experiments on liquid cesium at
308 K and the  commonly accepted value. After having accounted for
all corrections to the normalization of the neutron scattering
data, Bodensteiner {\it et al.}\cite{cs} inferred a (total)
incoherent cross section of 0.33 b instead of the literature value
of 0.22 b. Assuming that 0.22 b is indeed the correct value for
the incoherent cross-section, this would imply a magnetic
cross-section of 0.11 b, or $d\sigma^{magnetic}/d\Omega$ =
0.11/4$\pi$. Presumably, a collision would leave a cesium ion
temporarily with an iodine configuration ($S$=1/2, $L$=1, $J$= 3/2
and $g(LJS)$= 4/3), yielding $n$= 2.7\% (see Eq. \ref{para}).
Unfortunately, since uncorrected spectra at the smallest q-values
($q <$ 0.5 \AA$^{-1}$) were not published in this study\cite{cs},
we could not infer whether the supposed magnetic cross section
indeed corresponded to a quasi-elastic spectrum
characterized by a cage-diffusion linewidth.\\

From the current literature results, it is inconclusive whether
liquid potassium\cite{k,k2,bove2}, liquid rubidium\cite{ru,page},
or liquid
sodium\cite{na3,alblas,stangl,balucani2,pilgrim,scopigno} display
magnetic cross-sections. Either the data at low $q$ are not
accurate enough, or not enough details of the data correction
procedure have been given to test our thesis. Bearing in mind the
results for liquid cesium, the magnetic cross-section of $\sim$
0.1 b might just be too small to be observable in sodium
($\sigma_{inc}$ = 1.67 b) and rubidium ($\sigma_{inc}$ = 0.48 b).
However, the paramagnetic cross-section might have been observed
in liquid potassium ($\sigma_{inc}$ = 0.27 b) in a series of
quasi-elastic neutron scattering experiments\cite{k}. Cabrillo et
al.\cite{k} combined a high (energy) resolution study on liquid K
at 343 K with a lower resolution experiment to model the full
dynamic response of potassium down to small $q \sim$ 0.4
\AA$^{-1}$. Doing so, they were able to show that the
quasi-elastic component at small $q$ consisted of two
contributions, one corresponding to self-diffusion and one to a
process with a lifetime $\tau \sim $3 ps. Qualitatively, this is
similar to the observations for cage-diffusion in liquid mercury.
Unfortunately, the authors did not give the ratio between the
narrow and broad component, making it impossible to infer $n$ from
their data. In fact, the authors did not attribute this broad mode
to cage-diffusion. Instead, it was assumed to be part of the
coherent scattering contribution. The latter is inconsistent with
their modeling of the rest of the scattered intensity\cite{k},
which already completely exhausted the coherent sum-rule (Eq.
\ref{sq}). Given this, and given the very weak dependence of
$\tau$ on $q$ for $q <1.3$ \AA$^{-1}$, we believe that this broad
mode represents cage-diffusion. However, whether it is a
cage-diffusion
process combined with a fluctuating magnetic moment cannot be inferred from this study (as published).\\

Neutron scattering results for liquid lithium leave open the
possibility of a magnetic cross-section being present albeit that
the results are somewhat inaccurate owing to the large absorbtion
cross-section. For instance, Torcini {\it et al.}\cite{torcini}
report $S(q=0)$= 0.04 at 450 K, while the expected $S(q=0)$ from
the compressibility sum-rule is 0.03, thus indicating the presence
of a small magnetic cross-section. However, not all studies are in
agreement with these neutron scattering data (probably due to the
large absorbtion cross-section for neutrons). Therefore, we can
only give an estimated range for the fraction $n$ of ions with an
unpaired electron. Based on the work of Torcini {\it et
al.}\cite{torcini}, we find the fraction $n$ to be in the range
$0<n<1$\%,
for $S$=1/2, $L$=0, $J$=1/2 and $g(LJS)$=2.\\

In all, the alkali metals do not show unambiguous evidence for the
existence of the proposed magnetic cross-section. However, it is
interesting to note that small angle x-ray scattering experiments
on liquid lithium indicated the presence of an additional
cross-section\cite{burkel}, which the authors tentatively
attributed to increased correlation between the valence electrons.
The mechanism proposed in this paper would offer an explanation
for the observed\cite{burkel} increased correlation. Nonetheless,
the evidence for a collision-induced fluctuating moment in the
alkali metals is somewhat weak. Much better evidence for its
existence comes from scattering experiments on group 3 and 4
metals, which display an enhanced cross-section, similar to the results for liquid mercury.\\

\subsection{Group 3 and 4 metals}

Liquid lead is a good candidate to analyze for the possible presence of a magnetic cross-section since
Pb has a negligible incoherent cross-section; therefore, any significant scattering at small momentum
transfers (where the coherent cross-section is very small) is indicative of a paramagnetic signal.
Reijers {\it et al.}\cite{reijers} measured the static structure factor of liquid lead at 613 K under
ambient pressure (see Fig. \ref{pb}). From Eqs. \ref{main} and \ref{sq}, we find that the expected
neutron scattering intensity at small momentum transfers due to coherent scattering is given
by $\sigma_{coh}/4\pi S(q=0)$, with $S(q=0)$ = 0.009\cite{dubey} and $\sigma_{coh}$= 11.16 b.
The $S(q=0)$ extrapolated value from the liquid lead experiment is 0.07 (see Fig. \ref{pb}),
implying an additional neutron scattering intensity of 0.7/4$\pi$ b.  Using Eq. \ref{para}
with $S$= 1/2, $L$=2, $J$= 5/2 and $g(LJS)$= 1.2, the fraction $n$ of ions with an unpaired electron
is 9\%. Assuming the additional cross-section originates from s-electrons ($S$= 1/2, $L$=0, $J$= 1/2
and $g(LJS)$= 2), we find $n$= 38\% (See Table I).\\

Liquid aluminum also displays a paramagnetic cross-section.  Iqbal
{\it et al.}\cite{iqbal} performed a study on liquid aluminum at
936K (see Fig. \ref{al}). In this study on a liquid with
negligible incoherent cross-section, the authors normalized their
data to $S(q\rightarrow\infty)$= 1; however, the data had not been
corrected for multiple scattering effects, which can constitute a
major part of the scattering at small $q$. Based on the dimensions
of their cylindrical cell, we have calculated\cite{badyal,sears}
the multiple scattering contribution (dashed line in Fig.
\ref{al}) assuming the energy dependence of $S(q,E)$ to be given
by a Lorenzian line shape with half width determined by the
coefficient for self-diffusion ($D_s$= 0.4
\AA$^2$/ps\cite{gaskell}). After subtracting the multiple
scattering contribution and renormalizing the data accordingly, we
find that the neutron scattering data consistently lie above the
X-ray data\cite{xray} at small $q$, and that the neutron
scattering data do not appear to reach the $q \rightarrow 0$ limit
S(q=0) = 0.013\cite{alsound}. Since a paramagnetic contribution
represents a very small correction to X-ray scattering data, we
take the difference $\Delta S= 0.11$ between the neutron and X-ray
$S(q)$ measurements at $q < 1.5$ \AA$^{-1}$ as the strength of the
paramagnetic signal, i.e., $d\sigma^{magnetic}/d\Omega$= $\Delta
S$$\sigma_{coh}/4\pi$= 0.16/4$\pi$ b. This corresponds (Eq.
\ref{para}) to a fraction $n$= 4\% assuming the fluorine
electronic configuration for paramagnetic Al-ions; a sodium
configuration would correspond to $n$= 9\% (See Table I).\\

Another liquid metal for which we can verify the presence of an
additional component to the cross-section is liquid gallium.
Bellissent-Funel {\it et al.}\cite{bf} found in their experiments
on liquid Ga at 326 K and 959 K that the observed scattered
intensities were not consistent with the known values for
$\sigma_{inc}$ and $\sigma_{coh}$. Since both uncorrected and
corrected data were published in this study\cite{bf}, and since
every step of the data reduction procedure was clearly described,
we can infer a very accurate estimate of the paramagnetic
cross-section for Ga. Using the dimensions of the sample cell used
in the experiments\cite{bf}, we have calculated\cite{badyal,sears}
the multiple scattering contribution (see Fig. \ref{ga}). Taking
into account the $S(q=0)$ values and the fact that the magnetic
contribution will be absent at very large $q$, we find an
additional differential scattering cross-section of 0.88/4$\pi$ b
at 326 K and 0.78/4$\pi$ b at 959 K. Assuming this scattering to
originate from an unpaired electron with quantum numbers $S$= 1/2,
$L$=2, $J$= 5/2 and $g(LJS)$= 1.2, we find $n$ = 11.5\% at T= 326
K and $n$= 10.1\% at T=959 K. If we assume the scattering to
originate from a s-electron ($S$=1/2, L=0, $J$= 1/2 and $g(LJS)$=
2), we find $n$= 48\% and $n$= 42\%, respectively (see Table I).
Thus, gallium displays a large magnetic cross-section, but its
magnitude appears to be only weakly temperature dependent.

\section{Discussion}
The available neutron scattering data point overwhelmingly to the existence of short-lived magnetic
moments in non-magnetic liquid metals. These moments come in and out of existence on the same time scale
as the cage-diffusion motion, as observed in the quasi-elastic neutron scattering experiments on
liquid Hg\cite{liviaprl,badyal,bove}. The alkali metals show only a weak effect, but the effect is much
more pronounced in
mercury and in group 3 and 4 metals (see Table I).\\

The actual percentage of ions with unpaired electrons is more difficult to assess than establishing
that such ions with unpaired electrons exist. For instance, it is feasible that the unpaired electron
in liquid mercury is either an s-electron or a d-electron. The 6s shell in mercury has been drawn in
closer to the nucleus because of the relativistic contraction of the underlying shells, so it is
definitely conceivable in a liquid that the 6s shell can be completely filled (for some of the time at
least). In other words, the observed paramagnetic intensity could originate from a Hg$^{1+}$ or from
a Hg$^{3+}$-ion. (In liquid lead, it is in fact more likely that the paramagnetic contribution stems
from Pb$^{3+}$ than from Pb$^{5+}$-ions, given the prevalence of lead to form Pb$^{2+}$ in solids.)
Should this indeed be the case, then the electrical resistance in liquid mercury does not come solely
from electrons being scattered by ions, but also from electrons actually being captured by Hg-ions;
far from being unchanging, the Fermi-sea constantly changes in size while interchanging electrons
with the ions.\\

The phenomenon of the additional magnetic cross-section seems to have been mostly overlooked. However,
its implications on the interaction mechanisms in a liquid metal cannot be overlooked given the long
range of the magnetic dipole interaction and the ability of localized moments to polarize the surrounding
conduction electrons. In particular, it would be interesting to see how incorporation of paramagnetic
ions and their polarization capability into the interatomic potential used in MD simulations would alter
the characteristics of short-wavelength sound propagation.\\

Finally, this paramagnetic cross-section provides a means of studying the cage-diffusion mechanism at
small momentum transfers even in systems that do not exhibit an incoherent cross-section, such as lead
and aluminum. We are currently carrying out polarized neutron scattering experiments on liquid gallium
in order to verify that the observed additional cross-section is indeed magnetic in origin and to study
its temperature dependence close to the solidification transition.

\section{Acknowledgments}
We thank Dr. N. Iqbal for providing us with the diffraction data
on aluminum. Acknowledgement is made to the donors of the American
Chemical Society Petroleum Research Fund for support of this
research [ACS PRF 42615-G10].

\newpage
\begin{table}
\begin{tabular} { c c c c c c c c c}
Element & $T/T_{melting}$ & $S$ & $L$ & $J$ & $g(LJS)$&$\sigma_{magn}$ & $n$  & Ref. \\
&&&&&& [b] & [\%] & \\ \hline
Li & 1.03 & 1/2 & 0 & 1/2 & 2& 0-0.01 & 0-1 & \cite{torcini} \\
Al & 1.003 & 1/2 & 0 & 1/2 & 2& 0.16 & 9 & \cite{iqbal} \\
 &  & 1/2 & 1 & 3/2 & 1.33 &  & 4 &  \\
Ga & 1.075 & 1/2 & 0 & 1/2 & 2& 0.88 & 48 & \cite{bf} \\
 &  & 1/2 & 2 & 5/2 & 1.2 &  & 11.5 &  \\
 & 3.165 & 1/2 & 0 & 1/2 & 2& 0.78 & 42 & \cite{bf} \\
 &  & 1/2 & 2 & 5/2 & 1.2 &  & 10.1 &  \\
Cs & 1.022 & 1/2 & 1 & 3/2 & 1.33 & 0.11 & 2.7 & \cite{cs} \\
Hg & 1.25 & 1/2 & 0 & 1/2 & 2& 1.5 & 82 & \cite{badyal} \\
 &  & 1/2 & 2 & 5/2 & 1.2 &  & 19.5 &  \\
Pb & 1.021 & 1/2 & 0 & 1/2 & 2 & 0.7 & 38 & \cite{reijers} \\
 &  & 1/2 & 2 & 5/2 & 1.2 &  & 9 &  \\

\hline
\end{tabular}

\caption{The observed magnetic cross-section $\sigma_{magn}$ and the corresponding fraction $n$ of
ions with a magnetic moment, calculated for the most likely quantum numbers of the unpaired electron
using Eq. \ref{para}.}
\end{table}
\newpage

\begin{figure}
\caption[]
{Schematic representation of how cage diffusion can lead to short-lived magnetic moments. a) Snapshot
of a metallic liquid with ions showing completely filled shells. The Fermi-sea is not shown. b) On
close approach an electron is kicked out of an orbital. c) The resulting unpaired electron leads to
a local magnetic moment. d) This moment disappears again as the atoms move away from each other. }
\label{mechanism}
\end{figure}

\begin{figure}
\caption[] {The dynamic structure factor of liquid
mercury\cite{badyal} at small momentum transfer (solid circles)
and a vanadium reference sample (open circles) showing the
resolution of the neutron scattering spectrometer. The solid line
is a fit to two Lorentzian lines, taking the asymmetric
spectrometer resolution function into account. The bottom figure
is an enhancement of the top figure. One observes a sharp (in
energy, hence slow in time) central mode reflecting
self-diffusion, and a broad mode (dash-dotted curve) reflecting
the fast rattling motion of an atom inside the cage formed by its
neighbors. The intensity of this broad mode (clearly absent in the
vanadium spectra) was found to be larger\cite{badyal} by a factor
of 20 than could be expected from nuclear sum rules on the
scattering. Hence, the intensity was attributed to a paramagnetic
cross section, reflecting an unpaired d-electron on a time scale
determined by cage diffusion. (Figure reproduced from Ref.
\cite{badyal}).} \label{hg}
\end{figure}

\begin{figure}
\caption[] {The static structure factor of liquid lead as measured
by X-ray scattering data\cite{dzuxray} at 623 K (solid line) and
neutron scattering data\cite{reijers} at 613 K (stars). Note the
difference between the two data sets at small momentum transfer;
The X-ray scattering data approach $S(q=0)$ = 0.008 (open
diamond), while the neutron scattering data approach a constant
value well in excess of $S(q=0)$, indicative of a magnetic
contribution to the scattering} \label{pb}
\end{figure}

\begin{figure}
\caption[] {The static structure factor of liquid aluminum just
above the melting point as measured by neutron
scattering\cite{iqbal} (solid line) and X-ray
scattering\cite{xray} (stars). The difference between the two data
sets is considerably larger than the calculated multiple
scattering contribution to the neutron scattering data
(dashed-dotted curve). After correcting for these multiple
scattering effects, we find that the remaining difference between
the two data sets (solid circles and horizontal line) is only
weakly dependent on $q$, indicative of an incompletely filled
electronic shell with small radius. The data point at $q$=0 (open
diamond) is the compressibility limit taken from thermodynamic
data\cite{alsound}.} \label{al}
\end{figure}

\begin{figure}
\caption[] {The unnormalized static structure factor of liquid
gallium at two temperatures (solid line with stars) as measured by
neutron scattering\cite{bf}. The calculated incoherent
contribution is given by the dashed dotted lines, the sum of the
incoherent and multiple scattering contribution (see text) is
denoted by the solid lines. The difference (at small q-values)
between the experimental data points and the solid line is
ascribed to paramagnetic scattering.} \label{ga}
\end{figure}


\begin{thebibliography}{10}

\bibitem{liviaprl} L.E. Bove, F. Sacchetti, C. Petrillo, B. Dorner, F. Formisano, and F. Barocchi, Phys. Rev. Lett. {\bf 87}, 215504 (2001).
\bibitem{badyal}  Y.S. Badyal, U. Bafile, K. Miyazaki, I.M. de Schepper, and W. Montfrooij, Phys. Rev. {\bf E68}, 061208 (2003).
\bibitem{bove} L.E. Bove, F. Sacchetti, C. Petrillo, B. Dorner, F. Formisano, M. Sampoli, and F. Barocchi, Philos. Mag. {\bf 82}, 365 (2002); Journ. of Non-Cryst. Solids {\bf 307-310}, 842 (2002).

\bibitem{cs} T. Bodensteiner, Chr. Morkel, W. Glaeser, and B. Dorner, Phys. Rev. A {\bf 45}, 5709 (1992).
\bibitem{bove3} L.E. Bove, F. Sacchetti, C. Petrillo, and B. Dorner, Phys. Rev. Lett. {\bf 85}, 5352
(2000).

\bibitem{k} C. Cabrillo, F.J. Bermejo, M. Alvarez, P. Verkerk, A. Maira-Vidal, S.M. Bennington, and D. Martin, Phys. Rev. Lett. {\bf 89}, 075508 (2002).
\bibitem{k2} A.G. Novokov, M.N. Ivanovskii, V.V. Savostin, A.L. Shimkevich, O.V. Sobolev, and M.V. Zaezjev, J. Phys.: Condens. Matt. {\bf 8}, 3525 (1996).
\bibitem{bove2} L.E. Bove, B. Dorner, C. Petrillo, F. Sacchetti, and J.-B. Suck, Phys. Rev. B. {\bf 68}, 024208
(2003).

\bibitem{ru} J.R.D. Copley and J.M. Rowe, Phys. Rev. A {\bf 9}, 1656 (1974).
\bibitem{page} D.I. Page, P.A. Egelstaff, J.E. Enderby, and B.R. Wingfield, Phys. Lett. {\bf 29A}, 296
(1969).

\bibitem{na3} Chr. Morkel and W.-C. Pilgrim, Journ. Non-Cryst. Solids {\bf 312-314}, 128 (2002).
\bibitem{alblas} B.P. Alblas, W. van der Lugt, J. Dijkstra, W. Geertsma, and C. van Dijk, J. Phys. F.: Met. Phys. {\bf 13}, 2465
(1983).
\bibitem{stangl} A. Stangl, C. Morkel, U. Balucani, and A. Torcini, J. Non-Crys. Sol. {\bf 205-207}, 402
(1996).
\bibitem{balucani2} U. Balucani, A. Torcini, A. Stangl, and C. Morkel, Physica Scripta. {\bf T75}, 13
(1995).

\bibitem{torcini}  A. Torcini, U. Balucani, P.H.K. de Jong, and P. Verkerk, Phys. Rev. E. {\bf 51}, 3126 (1995).
\bibitem{burkel} H. Sinn and E. Burkel, J. Phys. Cond.Mat. {\bf 8}, 9369 (1996).
\bibitem{dejong} P.H.K. de Jong, P. Verkerk, and L.A. de Graaf, J. Phys.: Condens. Matt. {\bf 6}, 8391
(1994).

\bibitem{cocking} S.J. Cocking and P.A. Egelstaff, J. Phys. C. (Proc. Phys. Soc.). {\bf 1}, 507
(1968).
\bibitem{north} D.M. North, J.E. Enderby, and P.A. Egelstaff, J. Phys. C{\bf 1}, 784
(1968).
\bibitem{soderstrom} O. S\"oderstr\"om, J.R.D. Copley, J.-B. Suck, and B. Dorner, J. Phys. F.: Metal Phys. {\bf 10}, L151
(1980).
\bibitem{reijers} H.T.J. Reijers, W. van der Lugt, C. van Dijk, and M.-L. Saboungi, J. Phys.: Condens. Matt. {\bf 1}, 5229
(1989).

\bibitem{iqbal}  N. Iqbal, N.H. van Dijk, V.W.J. Verhoeven, W. Montfrooij, T. Hansen, L. Katgerman, and G.J. Kearley, Acta Materialia. {\bf 51}, 4497-4504
(2003).

\bibitem{bf}  M.C. Bellissent-Funel, P. Chieux, D. Levesque, and J.J. Weis, Phys. Rev. A. {\bf 39}, 6310 (1989).

\bibitem{fredrikze} H. Fredrikze, Phys. Rev. A. {\bf 36}, 2272
(1987).

\bibitem{egelstaff} P.A. Egelstaff, N.H. March, and N.C. McGill, Can. J. Phys. {\bf 52}, 1651
(1974).


\bibitem{tamura} K. Tamura, M. Inui, I. Nakaso, Y. Oh'ishi, K. Funakoshi, and W. Utsumi, J. Phys.: Condens. Matt. {\bf 10}, 11405
(1998).
\bibitem{tamura2} K. Tamura and S. Hosokawa, J. Phys.: Condens. Matt. {\bf 6}, A241
(1994).

\bibitem{huijben} M.J. Huijben and W. van der Lugt, J. Phys. F.: Met. Phys. {\bf 6}, L225
(1976).

\bibitem{pilgrim} W.-C. Pilgrim, S. Hosokawa, H. Saggau, H. Sinn, and E. Burkel, J. Non-Crys. Sol. {\bf 250-252}, 96
(1999).
\bibitem{scopigno} T. Scopigno, U. Balucani, G. Ruocco, and F. Sette, Phys. Rev. E. {\bf 65}, 031205
(2002).

\bibitem{li3} T. Scopigno, U. Balucani, G. Ruocco, and F. Sette, Phys. Rev. Lett. {\bf 85}, 4076 (2000).

\bibitem{dzuxray} M. Dzugutov, K.-E. Larsson, and I. Ebbsjo, Phys. Rev. A. {\bf 38}, 3609 (1988).

\bibitem{xray} IAMP database of SCM-LIQ, Tohoku University. Available at: http://www.tagen.tohoku.ac.jp/general/building/iamp/database/scm/LIQ/sq.html.

\bibitem{host} U. Balucani and M. Zoppi, {\it Dynamics of the Liquid State} (Clarendon Press, Oxford, 1994), and references therein.
\bibitem{cage} I.M. de Schepper, E.G.D. Cohen, and M.J. Zuilhof, Phys. Lett. {\bf 101A}, 399 (1984).
\bibitem{peter} E.G.D. Cohen, P. Westerhuijs, and I.M. de Schepper, Phys. Rev. Lett. {\bf 59}, 2872 (1987).
\bibitem{pusey} P.N. Pusey, H.N.W. Lekkerkerker, E.G.D. Cohen, and I.M. de Schepper, Physica A{\bf 164}, 12 (1990).
\bibitem{schepper} R. Verberg, I.M. de Schepper, and E.G.D. Cohen, Europhys. Lett. {\bf 48}, 397 (1999); Phys. Rev. E{\bf 61}, 2967 (2000).
\bibitem{squires}  G.L. Squires.  {\it Introduction to the Theory of Thermal Neutron Scattering}. New York: Dover, 1996.
\bibitem{sears} V.F. Sears, Adv. Phys. {\bf 24}, 1 (1975).
\bibitem{handbook} {\it Handbook of Chemistry and Physics}, 79$^{th}$ edition, CRC, 1998-1999.

\bibitem{dubey} G.S. Dubey, R. Bansal and K.N. Pathak, J. Phys. C: Solid St. Phys. {\bf 13}, 6119 (1980).
\bibitem{gaskell} T. Gaskell, J. Phys. F: Met. Phys. {\bf 16}, 381 (1986).
\bibitem{alsound} N.M. Keita and S. Steinemann, J. Phys. C: Solid State Phys. {\bf 11}, 4635 (1978).


\end{thebibliography}
\end{document}